\documentclass[conference]{IEEEtran}
\usepackage{pgf, adjustbox}

\usepackage{subfig}
\usepackage{csquotes}
\usepackage{hyperref}
\usepackage{balance}

\usepackage{graphicx}
\graphicspath{{img/}}

\usepackage{amsmath, amssymb, bbold, mathrsfs}
\usepackage[backend=bibtex, sorting=none]{biblatex}
\begin{document}
\title{Radio Imaging with Information Field Theory}
\author{
  \IEEEauthorblockN{
    Philipp Arras\IEEEauthorrefmark{1}\IEEEauthorrefmark{2},
    Jakob Knollm{\"u}ller\IEEEauthorrefmark{1},
    Henrik Junklewitz and
    Torsten A. En{\ss}lin\IEEEauthorrefmark{1} 
  }
  \IEEEauthorblockA{\IEEEauthorrefmark{1}Max-Planck Institute for Astrophysics,
    Garching, Germany}
  \IEEEauthorblockA{\IEEEauthorrefmark{2}Technical University of Munich, Munich,
  Germany}
}

\maketitle

\begin{abstract}
  Data from radio interferometers provide a substantial challenge for
  statisticians. It is incomplete, noise-dominated and originates from a
  non-trivial measurement process. The signal is not only corrupted by imperfect
  measurement devices but also from effects like fluctuations in the ionosphere
  that act as a distortion screen.
  In this paper we focus on the imaging part of data reduction in radio
  astronomy and present \textsc{RESOLVE}, a Bayesian imaging algorithm for radio
  interferometry in its new incarnation. It is formulated in the language of
  information field theory. Solely by algorithmic advances the inference could
  be sped up significantly and behaves noticeably more stable now. This is
  one more step towards a fully user-friendly version of \textsc{RESOLVE} which
  can be applied routinely by astronomers.
\end{abstract}

\IEEEpeerreviewmaketitle

\section{Introduction}
To explore the origins of our universe and to learn about physical laws on both
small and large scales telescopes of various kinds provide 
information.
An armada 
of telescopes including many radio telescopes all over the earth and in space collect data to be 
put into one consistent theoretical picture of our universe by
astrophysicists. 
Radio interferometers are of specific interest from a data
reductionist's point of view since they do not measure a direct image of the sky
as optical telescopes do. As a consequence radio interferometers provide only very incomplete
information about the patch of the sky they are looking at. These two factors
render the problem of radio imaging non-trivial and in order to obtain
high-quality images sophisticated statistical methods need to be developed and
applied. 

In this paper, we want to present the latest state of the art of
reducing data from radio interferometers with
the help of \emph{information field theory} (IFT) \cite{2013AIPC.1553..184E}.

IFT is a statistical field theory which enables
statisticians to solve complex Bayesian inference problems which involve fields.
A field is a physical quantity defined over a continuous space like a
three-dimensional density field or two-dimensional flux field. Treating these
fields as continuous objects IFT does not suffer from side-effects induced by
introducing a pixelation scheme right from the beginning. Moreover, a theory
formulated in the language of fields enables IFT statisticians to employ the
machinery having been developed by field theorists.

The algorithmic idea presented here is called \textsc{RESOLVE} (\textbf{R}adio
\textbf{E}xtended \textbf{SO}urces \textbf{L}ognormal decon\textbf{v}olution
\textbf{E}stimator) and was first presented in \cite{Resolve2016}. Since then
the inference machinery has evolved dramatically with subsequent speedups of a
factor of around 100.

This paper is organised as follows: In section~\ref{sec:measurement} the
measurement principle of radio interferometers is outlined.
Section~\ref{sec:ift} gives a quick introduction to information field theory
followed by section~\ref{sec:model} in which the Bayesian hierarchical model
used by RESOLVE is explained.
We conclude with an application on real data
in section~\ref{sec:application}.

\section{Measurement Process and Data in Radio Astronomy}
\label{sec:measurement}
Radio telescopes measure the electromagnetic sky in wave-lengths from $\lambda =
0.3\, \mathrm{mm}$ (lower limit of ALMA) to $30\, \mathrm{m}$ (upper limit of
LOFAR). This poses a serious problem. The angular resolution of a single-dish telescope
$\delta \theta$ scales with the wavelength $\lambda$ divided by the instrument
aperture $D$:
\begin{align*}
 \delta \theta = 1.22 \, \frac{\lambda}{D}.
\end{align*}

As an example consider $\lambda = 0.6 \, \mathrm{cm}$ and $\delta\theta = 0.1 \,
\mathrm{arcsec}$ which are typical values for the VLA. Then the size of the aperture would need to be approximately $15 \,
\mathrm{km}$ which is not feasible technically. Therefore, many radio telescopes
apply a different measurement principle.

Radio telescopes like VLA are in fact \emph{radio
  interferometers}. They consist of several antennas (a total number of 27 in the case of
the VLA). The electromagnetic radio wave which arrives at each antenna is
converted to a digital signal and sent to a central supercomputer, called
\emph{correlator}. As its name suggest, it correlates the signal of each antenna
with every other antenna in temporal windows of typically around $10\,\mathrm
s$. These correlation coefficients are called \emph{visibilities}. Each
visibility corresponds to the strength of excitation of a Fourier mode in image
space. The distance between two antennas is proportional to the spatial
frequency and the orientation of the antennas gives the orientation of the
Fourier mode.

All in all, the radio interferometric measurement process is modeled by the
\emph{Radio Interferometric Measurement Equation} (RIME, \cite{Smirnov:2011vp}):
\begin{align}
 \label{eq:rime}
 d_{pq} = \int I(l,m)e^{i(lu_p+mv_q)}\,dl\,dm + n_{pq}.
\end{align}
Put into words, the data is given by the Fourier transform of the flux
distribution $I(l,m)$ where $l$ and $m$ are the direction cosines of
the angular coordinates $\phi$ and $\theta$ on the sky. Please note that this formula is based on
several assumptions and simplifications. First, this version of the RIME is only
valid for narrow field of views since it assumes a flat sky. Second, it assumes
that all antennas are located at the same altitude. Third, it does not account for
different polarizations and assumes that the antennas simply measure Stokes $I$.
Finally and perhaps most importantly, it assumes that the data has
been perfectly calibrated for all possible instrumental and additional
measurement effects (e.g. receiver instabilties, ionispheric interference, $\ldots$).
In this paper we treat only radio imaging and build on
top of data which is calibrated by established algorithms. In other words, it is
assumed that the data is calibrated perfectly.

\section{Information Field Theory}
\label{sec:ift}
In a nutshell, IFT is information theory with
fields. It is a framework which uncovers the connection between statistical
field theory and Bayesian inference. Exploiting this connection enables us
to translate all knowledge physicists have gathered about statistical field theory
and thermodynamics to Bayesian inference.

The general idea is that given some finite data set $d$,
it is inferred how likely different realizations of the observed physical
field $s$ is. This is done with the help of Bayes theorem which combines the
likelihood $\mathcal P (d|s)$ with the prior knowledge $\mathcal P (s)$ and some
normalization constant $\mathcal P(d)$ into the posterior distribution $\mathcal
P(s|d)$:
\begin{align*}
\mathcal P (s|d) = \frac{\mathcal P(d|s) \mathcal P(s)}{\mathcal P(d)} = \frac{\mathcal P(s,d)}{\mathcal P(d)}.
\end{align*}
This can be rewritten as:
\begin{align*}
\mathcal P (s|d) = \frac{1}{\mathcal Z(d)} e^{-\mathcal H(s,d)},
\end{align*}
where $Z(d) := \int \mathcal Ds\, \mathcal P(s,d)$ and $\mathcal H(s,d) := -\log
\mathcal P(s,d)$. $\int\mathcal Ds$ is the path integral which is defined as the
continuum limit of the product of integrals over every pixel $\int \prod_i ds_i$. For
details on that refer to \cite{2013AIPC.1553..184E, wiki:ift}.

The above formula is well-known in statistical physics and inspires
us to call $\mathcal H$ the \emph{information Hamiltonian}. In order to obtain the
maximum a-posterior estimate (MAP) of $s$ one has to minimize $\mathcal H$
with respect to $s$ because the exponential is a monotonic increasing function.
Since the information Hamiltonian is given by
\begin{align*}
  \mathcal H(s,d) = \mathcal H(d|s) + \mathcal H(s),
\end{align*}
it knows both about the measurement process via the likelihood term $\mathcal
H(d|s)$ and about the prior knowledge via $\mathcal H(s)$. Please note that
additional constants in $s$ can be dropped from $\mathcal H(s,d)$ since they
only change the normalization of the posterior but not its shape. This will be
indicated by \enquote{$\simeq$}.

As an illustrative example, let us re-derived the famous Wiener filter
\cite{wiener1949extrapolation}. Suppose we observe a noisy random process with
known stationary signal and noise spectra and additive noise. More precisely,
suppose we are given some measurement data $d$ described by the following
measurement equation:
\begin{align}
  \label{eq:measurement}
 d = Rs + n,
\end{align}
where $d$ is a finite-dimensional vector, $s$ is the unknown signal field and
$n$ the additive noise. $s$ and $n$ are assumed to be zero-centered Gaussian random fields
drawn from $\mathscr G(s, S)$ and $\mathscr G(n,N)$, respectively, where the
covariances $S$ and $N$ are known. $R$, the linear \emph{Response operator},
models the measurement device and is also known. It maps the signal $s$ defined
over a continuous domain to a finite data vector $d$. Note that
equation~\eqref{eq:rime}, the RIME, is of that form. Also note that in this
specific case the response operator $R$ contains a Fourier transform.

Let us compute the posterior distribution or equivalently the information
Hamiltonian for this problem. The likelihood $\mathcal P(d|s)$ is essentially
given by equation~\eqref{eq:measurement}:
\begin{align*}
 \mathcal P(d|s, n) = \delta (d - (Rs + n)).
\end{align*}
Then marginalize over the noise field:
\begin{align*}
 \mathcal P(d|s) = \int \mathcal D n \, \mathcal P(d|s, n)\mathcal P(n) = \mathscr G(d-Rs, N).
\end{align*}
Combining this with the prior probability $\mathcal P(s) = \mathscr G (s,S)$ and
taking the negative logarithm
gives the information Hamiltonian:
\begin{align}
  \label{eq:infHamiltonian}
  \begin{split}
    \mathcal H (s,d) =&\, \tfrac12 (d-Rs)^\dagger N^{-1} (d-Rs) +\tfrac12 s^\dagger S^{-1} s\\
    &\quad-\tfrac12 \log |2\pi N| -\tfrac12 \log |2\pi S|,
  \end{split}
\end{align}
where $\cdot^\dagger$ denotes transposition and element-wise complex conjugation
of a matrix or a vector. The above expression is a second order polynomial and
the square in $s$ can be completed:
\begin{align*}
 \mathcal H(s,d) \simeq \tfrac12 (s-m)^\dagger D^{-1} (s-m),
\end{align*}
where $m=Dj$, $j = R^\dagger N^{-1} d$ and $D^{-1} = S^{-1} + R^\dagger N^{-1}
R$. In other words, the posterior probability distribution is
\begin{align*}
\mathcal P(s|d) = \mathscr G(s-m, D)
\end{align*}
where $m$ is called the Wiener filter solution.

In this fashion the Wiener filter turns out to be the simplest filter which can
be build within the framework of IFT. Note that already here one of IFT's strength
becomes apparent: Pixelation schemes have not appeared yet. This is
a general feature of IFT. The theory is formulated with fields (which infinitely
many degrees of freedom which are not pixelated yet). Only when the filter is implemented on the computer
the fields become discretised. To this end the Python package NIFTy
\cite{nifty,NIFTY1,NIFTY3}
provides customized functionality to implement IFT algorithms. It even enables the user to easily switch between
different pixelation schemes.

\begin{figure*}[t!]
\centering
\subfloat[Posterior mean $m$ (logarithmic brightness).]{\includegraphics[width=2.5in]{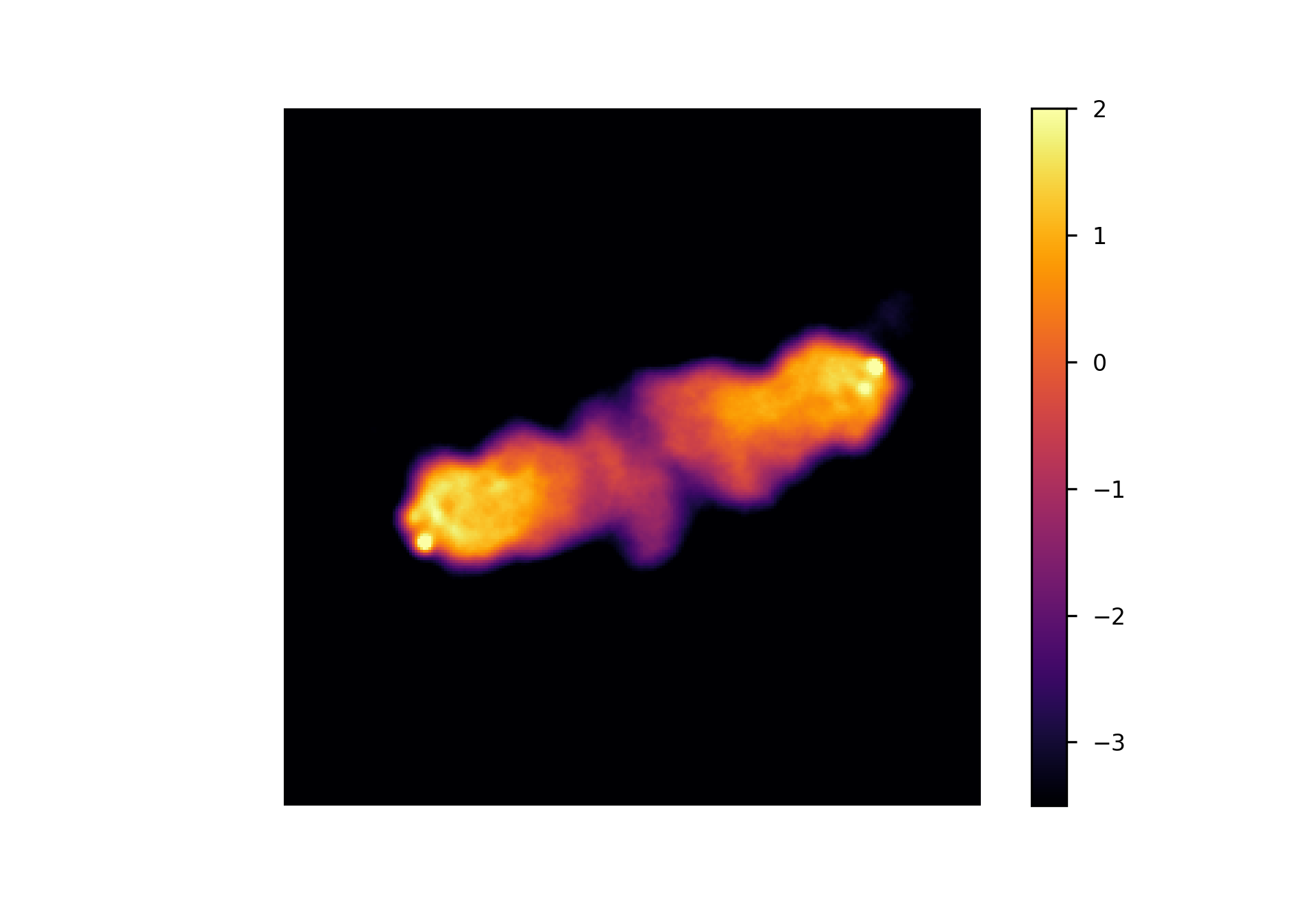}%
\label{fig:cygnusAmap}}
\hfil
\subfloat[Relative error on $m$.]{\includegraphics[width=2.5in]{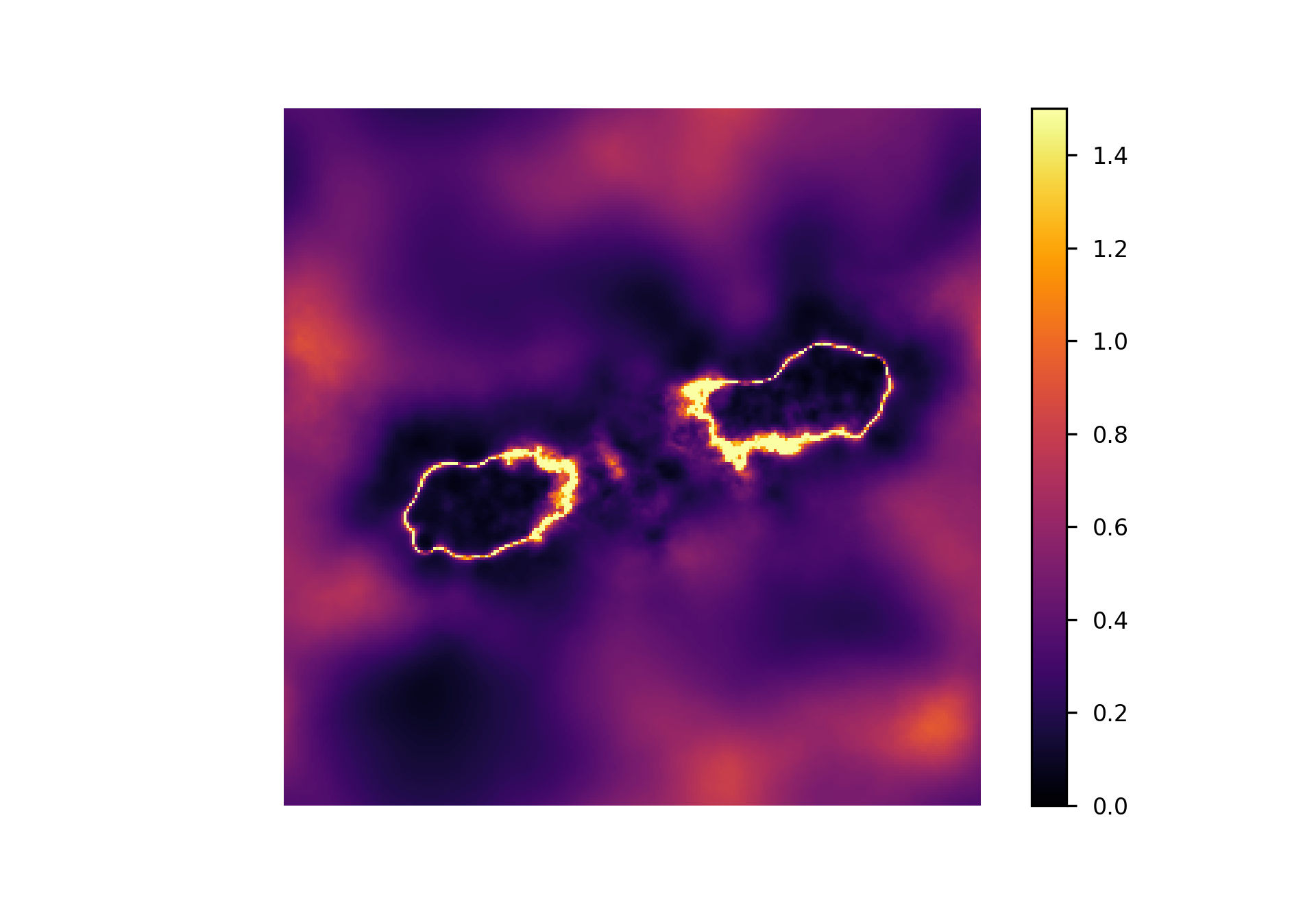}%
\label{fig:cygnusAuncertaintymap}}
\caption{Exemplary application of RESOLVE on real data which was taken in 2003
  by the VLA of the source 3C405 also known as Cygnus A.}
\label{fig:cygnusA}
\end{figure*}

\section{IFT Model for Radio Interferometers}
\label{sec:model}

In radio interferometry, the situation is somewhat more difficult than the Wiener
filter scenario discussed so far: First, the radio sky cannot be sensibly modeled by a Gaussian random
process since electromagnetic flux is always positive and varies on many
different orders of magnitude: a radio source typically is many magnitudes brighter than 
the surrounding background flux. Second, we do not know the signal covariances $S$
of the brightness distribution on the sky. Therefore, we need to infer it as
well. And finally, the noise covariance provided by the telescope might not be
entirely correct. Radio frequency interference or calibration errors might
enhance the error bars on the data significantly. Therefore, the noise level of
each data point needs to be inferred as well. The underlying assumptions and priors of
the following calculations are:
\begin{enumerate}
\item The sky obeys log-normal statistics, i.e. the measurement can be written
  as:
  \begin{align*}
    d = R e^s + n, 
  \end{align*}
  where $s$ is a Gaussian field again and $R$ is the linear response operator
  which maps the sky field onto visibilities.\footnote{Here and in the
    following, exponentials of vectors are understood to be taken element-wise.}
  This is the proper choice since it enforces positivity of the flux field and
  can easily vary on different scales.
 \item $s$ is drawn from a probability distribution describing a isotropic and
   homogeneous process.
 \item Power spectra of $s$ preferentially follow a power law. In other
   words, curvature on double-logarithmic scale in the power spectrum shall be
   punished in the inference.
 \item The noise covariance matrix is diagonal: $N= \widehat{e^{\eta}}$, where
   $\eta$ is a vector whose entries are the logarithms of the variance of every
   data point.\footnote{The hat operator $\widehat{e^\eta}$ denotes the diagonal
     operator with the vector $e^\eta$ on its diagonal.}
 \item Large noise covariances are punished by an Inverse-Gamma prior on $\eta$.
 \item The posterior probability distribution can be approximated by $\tilde{\mathcal P}
   (s, \tau, \eta | d) = \mathscr G (\xi-t, \Xi)\, \delta(\tau - \tau^*)\,
   \delta (\eta - \eta^*)$, where $\tau$ is the logarithm of the power spectrum
   and $\Xi$ is the posterior covariance of the map estimation.
\end{enumerate}

For starters let us introduce some notation.
Because $s$ is drawn from an
isotropic and homogeneous probability distribution the Wiener-Khinchin theorem
\cite{wiener1930generalized} implies that $S$ is diagonal in Fourier space and
its diagonal is given by a power spectrum $p(k)$:
\begin{align*}
 S_{\vec k \vec k'}  = (2\pi)^2 \delta (\vec k - \vec k') \, p(|\vec k|).
\end{align*}
The power spectrum is a positive function, thus we can apply the same trick as
for the sky map. Define:
\begin{align*}
 p(|\vec k|) = e^{\tau (|\vec k|)}
\end{align*}
For convenience define a projection operator $\mathbb P$ which sums all values
of a field $b$ in harmonic space which lie in one bin in the power spectrum:
\begin{align*}
 b_{\vec k} = \mathbb P_{\vec k \kappa} a_{\kappa} = \frac1{\rho_k}\int_{|\vec k| = \kappa} p_{\kappa},
\end{align*}
where $\rho_k$ is the bin volume. Defining $\mathcal F$ to be the Fourier
transform mapping from harmonic space to signal space, the signal prior
covariance $S$ can be expressed as:
\begin{align*}
 S = \mathcal F \left( \widehat{\mathbb P^\dagger e^\tau} \right) \mathcal F^\dagger.
\end{align*}
Finally, we split the field $s$ into two parts in harmonic space: $s = \mathcal
F(A_\tau\,\xi)$. $\xi$ is a white Gaussian random field, i.e. it has the covariance
matrix $\mathbb{1}$, and $A_\tau = \mathbb P^\dagger \sqrt{e^\tau}$, i.e. it contains
all information coming from the power spectrum.

With the above notation it is now possible to write down all Hamiltonians we
need for the reconstruction. The Hamiltonian which is to be minimized for the
$\xi$ reconstruction is computed analogously to \eqref{eq:infHamiltonian}:
\begin{align*}
 \mathcal H(\xi, d|\tau, \eta) \simeq \tfrac12 (d-R e^{\mathcal F(A_\tau\xi)})^\dagger \widehat{e^{-\eta}} (d-Re^{\mathcal F(A_\tau\xi)}) + \tfrac12 \xi^\dagger \xi.
\end{align*}
Since it will be needed later, the curvature of the above Hamiltonian is to be
computed:
\begin{align*}
  \Xi := \frac{\delta^2 \mathcal H(\xi, d|\tau, \eta)}{\delta\xi\,\delta\xi^\dagger}
          &= A_\tau^\dagger (e^s)^\dagger R^\dagger N^{-1} R e^s A_\tau+ \mathbb 1\\
          &\quad\quad\quad\quad-(d-Re^s)^\dagger N^{-1} R e^s A_\tau A_\tau.
\end{align*}
The last term is not necessarily positive definite which is not allowed for a
covariance operator\footnote{Note that the curvature of the information
  Hamiltonian is at the same time used as an approximative covariance of the posterior.}. However, this term is small in the vicinity
of the minimum because it contains the residual $d-Re^s$. Therefore, it
is dropped right from the beginning. 

The Hamiltonian for the power spectrum reconstruction has a very similar
structure: The likelihood is accompanied by the prior. Here, we choose a
smoothness prior on double-logarithmic scale. $\Delta$ is the Laplace operator acting
on logarithmic scale $y=\log k$:
\begin{align*}
  \begin{split}
    \mathcal H(\tau, d| \xi, \eta) \simeq &\,\tfrac12 (d-R e^{\mathcal F(A_\tau\xi)})^\dagger \widehat{e^{-\eta}} (d-Re^{\mathcal F(A_\tau\xi)})\\
    &\quad + \tfrac1{2\sigma^2} \tau^\dagger \Delta^\dagger \Delta\tau.
  \end{split}
\end{align*}
The parameter $\sigma$ controls the strength of the smoothness prior.

The Hamiltonian for the noise covariance estimation has again the same structure
except for the prior: Here, an Inverse-Gamma prior is employed:
\begin{align*}
 \mathcal H(\eta, d|\xi, \tau) \simeq &\,\tfrac12 (d-R e^{\mathcal F(A_\tau\xi)})^\dagger \widehat{e^{-\eta}} (d-Re^{\mathcal F(A_\tau\xi)})\\
 &\quad + \eta^\dagger (\alpha -1) + q^\dagger e^{-\eta} +\tfrac12 \mathbb 1^\dagger \eta.
\end{align*}
Note that the last term originates from the term $-\tfrac12 \log |2\pi N|$ in
\eqref{eq:infHamiltonian}.

In order to compute an estimate for the posterior $\tau^*$ and $\eta^*$, the
deviation between the correct posterior probability and the approximate one
needs to be minimized. The metric of choice to compare probability
distributions is the Kullbach-Leibler divergence:
\begin{align*}
\mathcal D_{\mathrm{KL}} (\tilde{\mathcal P} (\xi, \tau, \eta | d)\, \| \,\mathcal P (\xi, \tau,\eta | d)) = \int \mathcal D\xi \,\mathcal D\tau\,\mathcal D\eta\, \tilde{\mathcal P} \log \frac{\tilde{\mathcal P}}{\mathcal P}.
\end{align*}
The posterior shall be approximated by the distribution:
\begin{align*}
\tilde{\mathcal P}(s, \tau, \eta | d) =
\mathscr G (\xi -t, \Xi)\,\delta(\tau-\tau^*)\,\delta(\eta-\eta^*).
\end{align*}
The integrals over $\tau$ and $\eta$ simply collapse due to the
$\delta$-distributions. What remains are two objective function, one for the
power spectrum and one for the noise covariance estimation:
\begin{align*}
  \mathcal D_{\mathrm{KL}, \tau} &= \left\langle \tfrac12 (d-R e^{\mathcal F(A_\tau\xi)})^\dagger \widehat{e^{-\eta}} (d-Re^{\mathcal F(A_\tau\xi)}) \right\rangle_{\mathscr G(\xi - t, \Xi)}\\
  &\quad\quad\quad+ \frac1{2\sigma^2}\tau^\dagger\Delta^\dagger\Delta\tau,\\
  \mathcal D_{\mathrm{KL}, \eta} &= \left\langle \tfrac12 (d-R e^{\mathcal F(A_\tau\xi)})^\dagger \widehat{e^{-\eta}} (d-Re^{\mathcal F(A_\tau\xi)}) \right\rangle_{\mathscr G(\xi - t, \Xi)}\\
  &\quad\quad\quad+ (\alpha -1)^\dagger \eta + q^\dagger e^{-\eta}+\tfrac12 \mathbb 1^\dagger \eta.
\end{align*}
The expectation value $\langle \ldots \rangle_{\mathscr G (\xi-t, \Xi)}$ can be
computed by sampling from $\mathscr G(\xi-t, \Xi)$. For details on that refer to
\cite{2017arXiv171102955K}.

All in all, the complete inference algorithm for applying IFT to radio
interferometric data has been derived. The free parameters of the machinery are:
the strength of the smoothness prior on the power spectrum $\sigma$ and the
shape of the Inverse-Gamma prior on the noise covariance estimation $\alpha$ and
$q$.

\section{Application}
\label{sec:application}
Finally, let us apply the above derived Bayesian inference algorithm to real
data. To this end, let us take a VLA measurement set of Cygnus A from 2003. It
has a total integration time of 49100 seconds. Since we deal only with
single-band imaging in this paper, let us take one channel centered at
$327.5$~MHz with a bandwidth of $2.8$~Mhz. As prior settings we choose an
uninformative flat Inverse-Gamma prior for the noise ($q=10^{-5}$, $\alpha = 2$)
and $\sigma = 1$ for the smoothness prior on the power spectrum.

\begin{figure}[!t]
  \centering
  \input{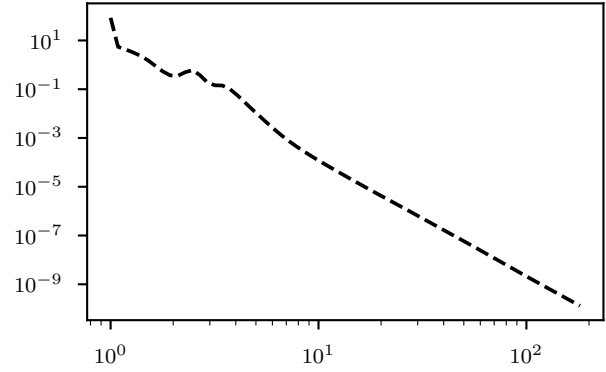} 
  \caption{Power spectrum of Cygnus A reconstruction.}
  \label{fig:cygnusApowerspectrum}
\end{figure}

The main result is presented in Figure~\ref{fig:cygnusA}. It shows the mean of
the Gaussian which approximates the sky part of the posterior: $\mathscr G
(s-m,D)$. Note that the figure shows the logarithmic flux. What singles out
RESOLVE from many other imaging algorithms is its ability to provide an
uncertainty map. It is depicted on the right-hand side on
Figure~\ref{fig:cygnusA}. Additional to the sky model the algorithm learns the
power spectrum $e^\tau$ as well. It is shown in Figure~\ref{fig:cygnusApowerspectrum}.
Note that it does not possess much curvature on log-log scale as was expected by
the Laplace prior on $\tau$.

\begin{figure*}[t!]
\centering
\subfloat[Error bars provided by telescope.]{\includegraphics[width=2.5in]{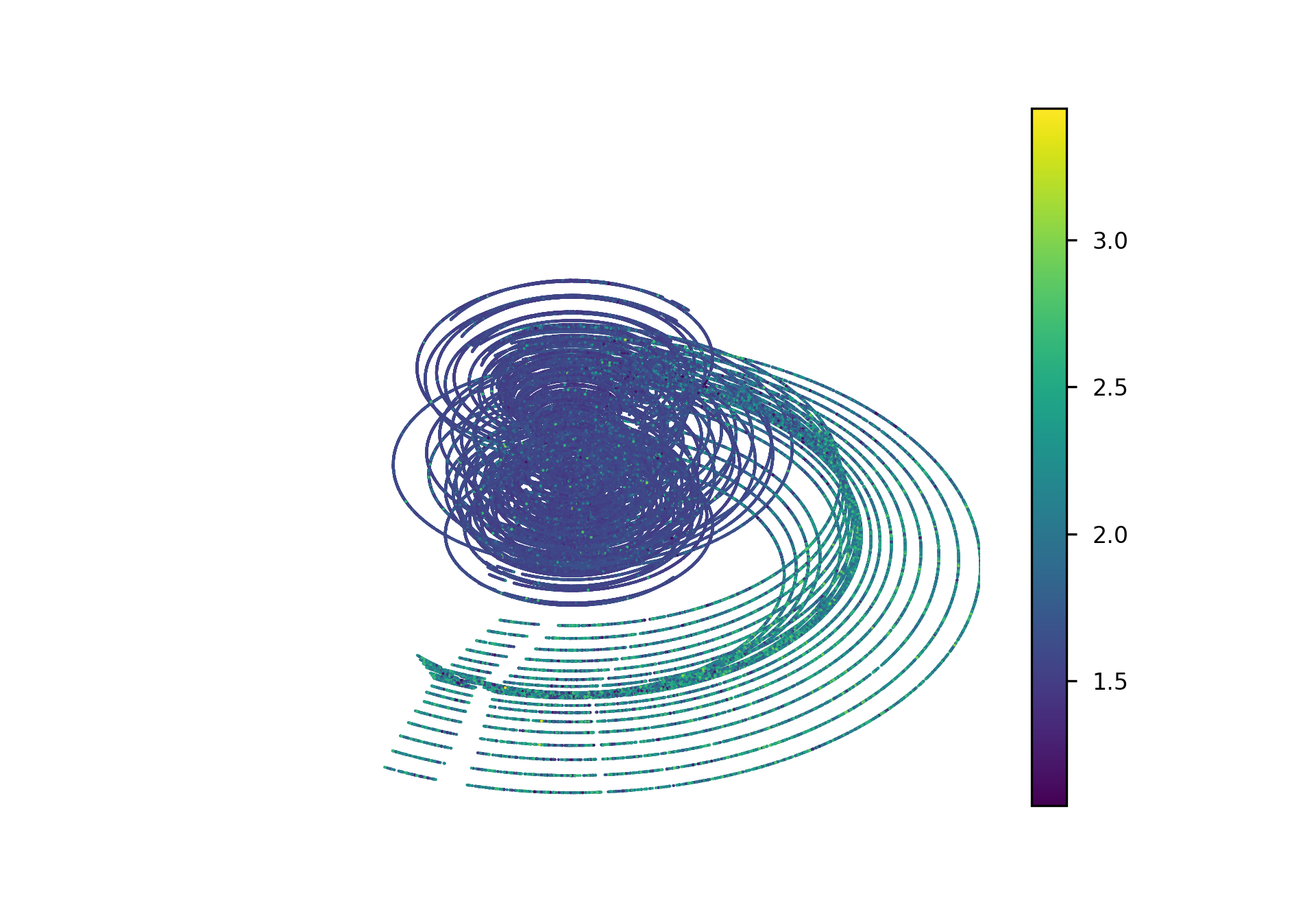}%
  \label{fig:cygnusAuv1}}
\hfil
\subfloat[Error bars provided by RESOVLE.]{\includegraphics[width=2.5in]{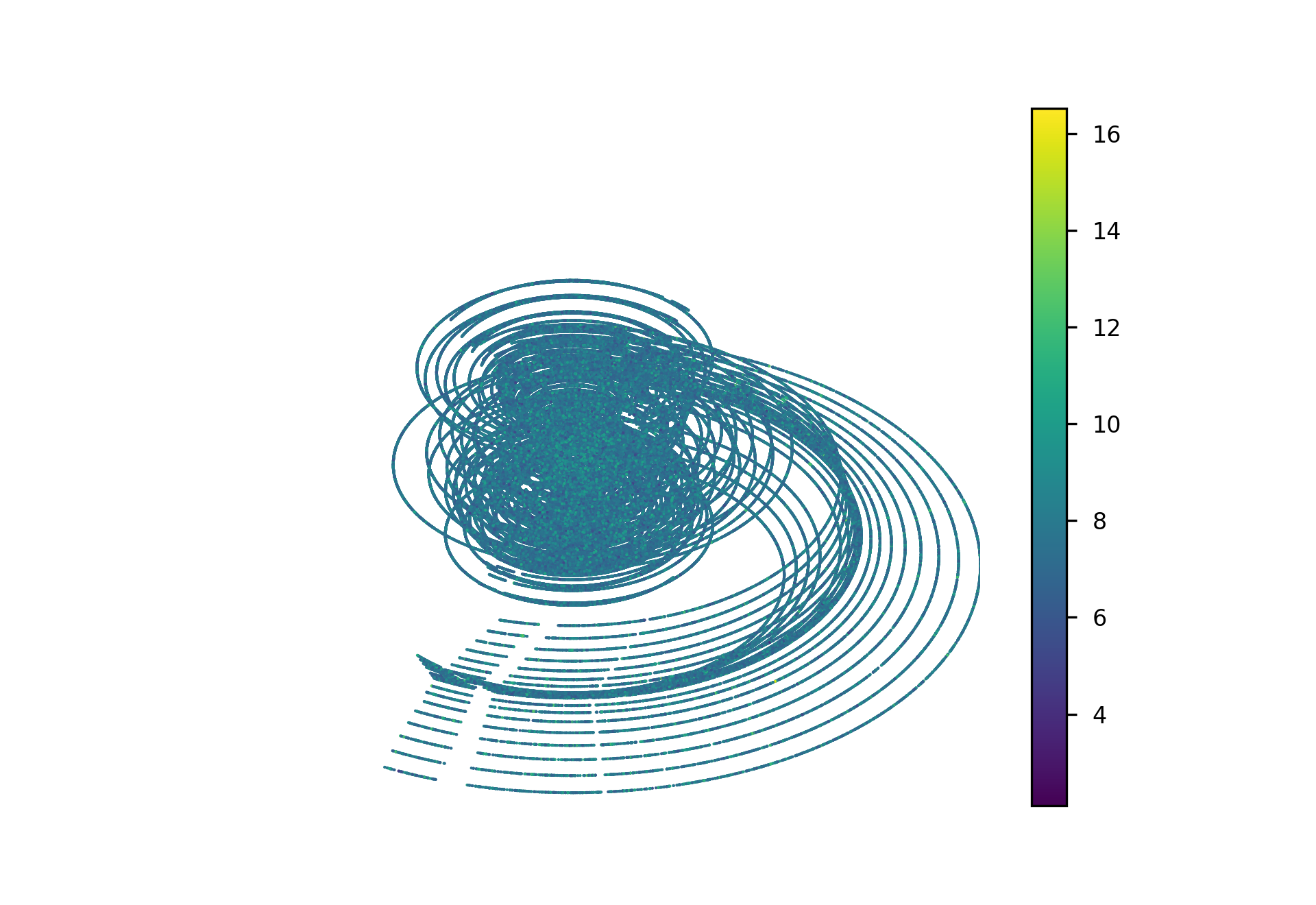}%
  \label{fig:cygnusAuv2}}
  \caption{Comparison of errorbars provided by the telescope and by RESOLVE. In
    the figures the logarithm of the variance of the data points is depicted.}
  \label{fig:cygnusAuv}
\end{figure*}

Finally, RESOLVE provides errorbars on the data points (see
Figure~\ref{fig:cygnusAuv}). It is apparent the RESOLVE's error bars are five
orders of magnitude bigger than the errorbars which are provided by the
telescope.

The reconstruction was run on an Intel Core~i5-4258U CPU using 300~MB main
memory. The resolution of the reconstruction is $256^2$~pixels for the sky model
and 32~pixels in the power spectrum. The response operator $R$ which
incorporates a nonequispaced fast Fourier transform was implemented by employing
the NFFT library which provides OpenMP parallelization \cite{keiner2009using}.

The reconstruction including the analysis of the posterior statistics took
approximately two hours of wall time.

\section{Conclusion}
\label{sec:conclusion}
In this paper \textsc{RESOLVE} in its new incarnation was presented for the
first time. Minimizing the Hamiltonian with respect to the map and the
KL-divergence with respect to the power spectrum and the noise level provide a
major speed-up. Also, the noise level of each data point was learned
simultaneously with the map reconstruction for the first time. The main insights
are:
\begin{itemize}
\item \textsc{RESOLVE}'s noise estimation suggests a much higher noise level
  compared to the noise level which comes with the data set. This might be
  rooted in calibration artifacts which \textsc{RESOLVE} detects and puts into
  the noise.
\item The migration from a simple fix-point iteration to minimization of
  Hamiltonian and KL-divergences was successful and is a big step forward
  towards an easy-to-use version of \textsc{RESOLVE} which can be shipped to a
  broad range of end-users.
\end{itemize}
The apparent next step towards a fully-integrated IFT radio data reconstruction
pipeline is to include the calibration into the IFT inference. Other possible
future work is to develop a fancier radio response function which can deal with
wide-field images and to include point source reconstructions in the spirit of
\cite{2018arXiv180202013P}.

\section*{Acknowledgment}

The authors would like to thank Rick Perley for the calibrated Cygnus A data and
Landman Bester, Philipp Frank, Reimar Leike, Martin Reinecke, Oleg Smirnov and
R{\"u}diger Westermann for numerous helpful discussions.

We acknowledge financial support by the German Federal Ministry of Education and
Research (BMBF) under grant 05A17PB1 (Verbundprojekt D-MeerKAT).

\balance
\printbibliography

\end{document}